\documentclass{JAC2003}
\usepackage{graphicx, subfigure,amsmath}
\usepackage{booktabs}

\begin{document}
\title{Beam Coupling Impedance Simulations of the LHC TCTP Collimators}
\author{H. Day\thanks{hugo.day@hep.manchester.ac.uk}$^{\dagger\ddagger\star}$, F. Caspers$^{\dagger}$, A. Dallocchio$^{\dagger}$, L. Gentini$^{\dagger}$, A. Grudiev$^{\dagger}$, E. Metral$^{\dagger}$, B. Salvant$^{\dagger}$, R.M. Jones$^{\ddagger\star}$\\
$\dagger$ CERN, Switzerland \\
$\ddagger$ School of Physics and Astronomy, The University of Manchester, Manchester, UK \\
$\star$ Cockcroft Institute, Daresbury, UK \\}

\maketitle

\begin{abstract}
As part of an upgrade to the LHC collimation system, 8 TCTP and 1 TCSG collimators are proposed to replace existing collimators in the collimation system. In an effort to review all equipment placed in the accelerator complex for potential side effects due to collective effects and beam-equipment interactions, beam coupling impedance simulations are carried out in both the time-domain and frequency-domain of the full TCTP design. Particular attention is paid to trapped modes that may induce beam instabilities and beam-induced heating due to cavity modes of the device.
\end{abstract}


\section{Introduction}

The TCTP is a tertiary collimator with built-in beam postion monitors (BPMs). It is proposed to replace 8 existing phase 1 TCT collimators with 8 of these collimators\cite{dallo_ipac2011}. Asides from the addition of embedded BPMs, the other major new feature of these collimators is the implementation of new system of beam impedance mitigation. 

\begin{figure}
\subfigure[]{
\includegraphics[width=0.35\textwidth]{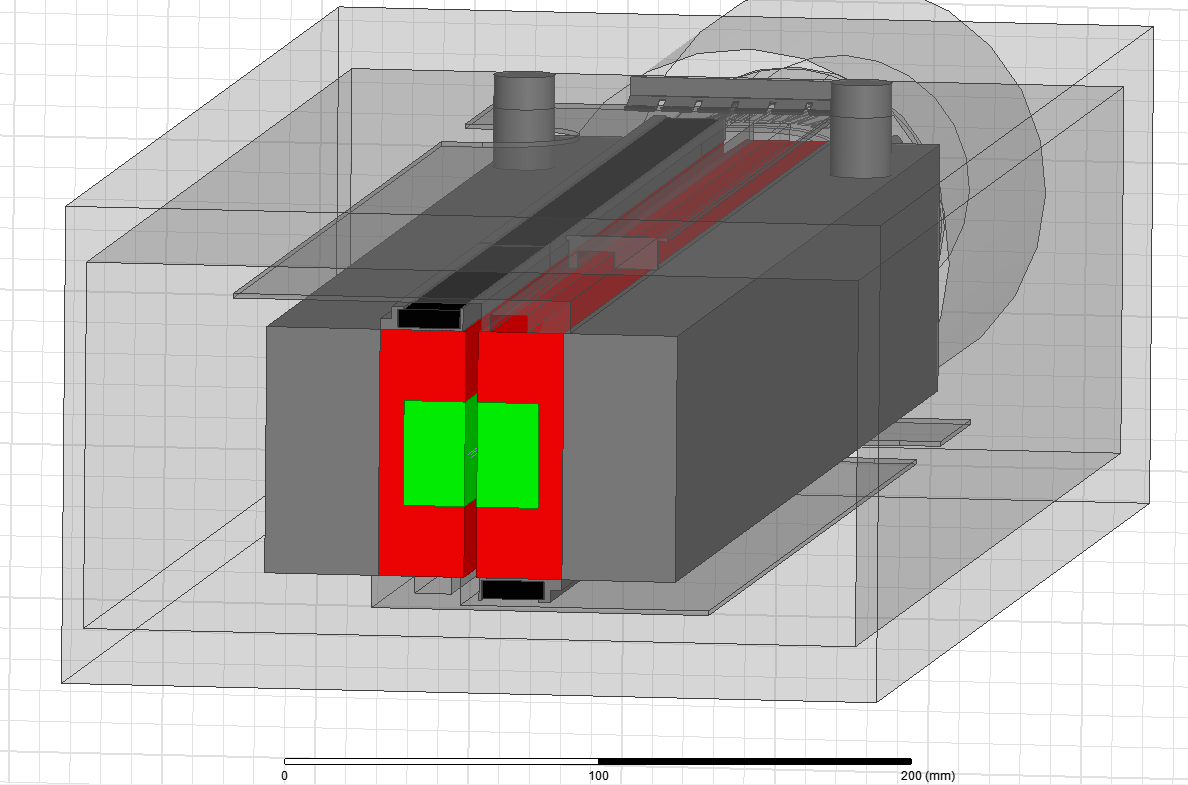}
\label{fig:tctp-ferrite}
}
\subfigure[]{
\includegraphics[width=0.35\textwidth]{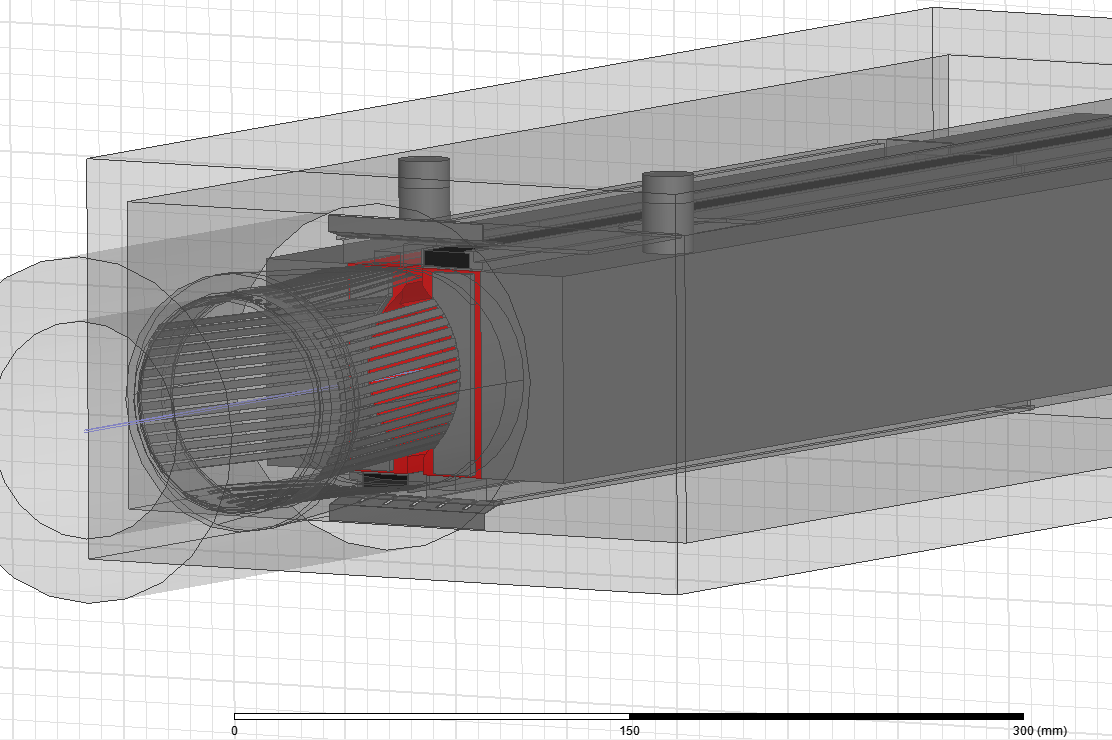}
\label{fig:tctp-fingers}
}
\label{fig:tctp_whole}
\caption{The design of the TCTP structure illustrating \ref{fig:tctp-ferrite} the ferrite damping circuit and \ref{fig:tctp-fingers} the RF fingers from the vacuum to the collimator jaw. Ferrite is in black, copper in red, tungsten in green and stainless steel in grey.}
\end{figure}

For the phase 1 collimators, the vacuum tank was screened from the beam by the use of sliding RF contacts; for the phase 2 design this will be replaced with a beam screen and ferrite damping tiles (see Fig.~\ref{fig:tctp-ferrite}). The phase 1 design is intended to screen the volume of the vacuum tank from the beam, thus causing the cavity modes to occur at high frequencies where the beam power spectrum is low. On the other hand, the phase 2 design allows the vacuum tank to be seen by the beam, but uses carefully placed ferrite tiles to significantly reduce the Q of the resulting cavity modes.

To verify the effectiveness of this new design and provide comparison due to the phase 1 design, we have carried out electromagnetic simulations in both the time and frequency domain to identify cavity modes, quantify the reduction in beam coupling impedance and localise the power loss by the beam to identify any possible issues with beam induced heating. Evaluation of the cavity modes of the phase 1 design can be found in Ref.~\cite{phase1sec}.

\section{Beam Coupling Impedance Simulations}

The impedance simulations of the TCTP were carried out using both time domain and frequency domain codes. For the time domain simulations, CST Particle Studios' \cite{cst-cite} wakefield solver was used to obtain the wakefield of a gaussian bunch with a sigma of 60mm of the full TCTP structure, modelled in a hexahedral mesh totalling 28 million mesh cells. A FFT algorithm is then used to obtain the beam coupling impedance. In the frequency domain eigenmode simulations have been carried out using HFSS \cite{hfss-cite} to identify the cavity modes of the structure. The resonant frequencies, shunt impedance and Q-factor of these resonances can then be identified. Due to computational constraints, we simulate half of the structure of the TCTP (see Fig.~1) changing the z-boundary between E- and H- boundaries to obtain two orthonormal families of eigenmodes. Mesh counts vary between 150,000-300,000 depending on the frequency of the solved mode.

It is possible to calculate the impedance of a cavity resonance from \cite{cav-res}

\begin{equation}
Z_{bb}\left( \omega \right) = \frac{R_{s}}{1+jQ\left( \frac{\omega}{\omega_{res}} -  \frac{\omega_{res}}{\omega} \right)}
\end{equation}

where $R_{s}$ is the shunt impedance, $\omega$ is the frequency, $\omega_{res} = 2\pi{}f_{res}$, $f_{res}$ is the eigenfrequency of the resonance, and Q is the quality factor of the resonance.

\begin{figure}
\includegraphics[width=0.5\textwidth]{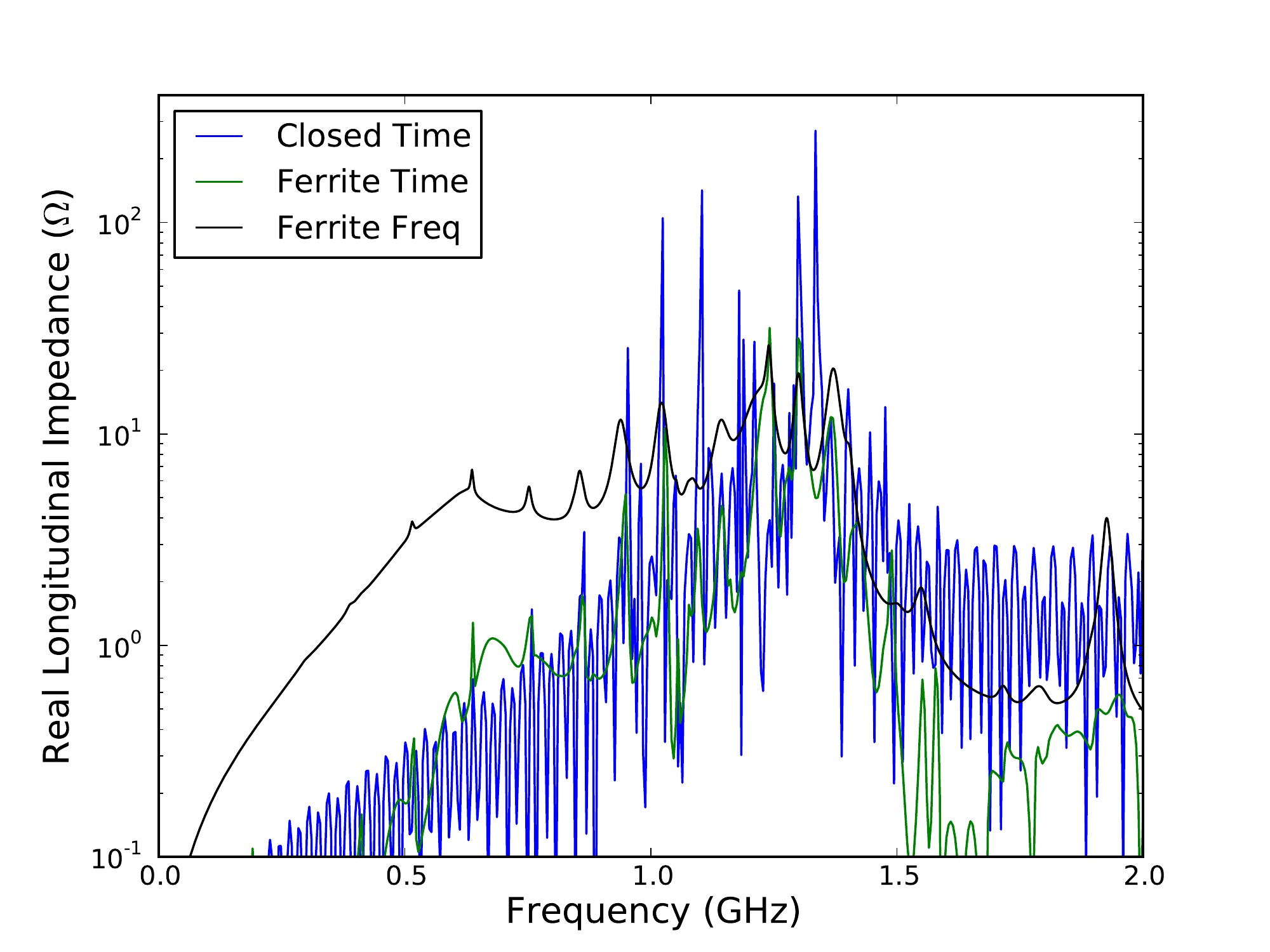}
\label{fig:bci_comparison}
\caption{The beam coupling impedance of the TCTP from both time domain and frequency domain simulations, compared to that of a simple closed collimator (of phase 1 design). Note the resonances is the closed structure are due to a truncation of the simulated wakepotential. This is due to computational limitations of the simulations}
\end{figure}

The time domain and frequency domain results for the broadband impedance are shown in Fig.~2. It can be seen that the frequencies of the peaks match well, however the height of the peaks do not match well and the broadband impedance is significantly different. This is due to the difference in materials definitions in the two simulations; realistic materials were used in the frequency domain simulations (i.e. vacuum tank as stainless steel) whereas a perfect conductor (PEC) was used for non-ferrite structures in the time domain simulations (i.e. vacuum tank as a perfect conductor), and due to a truncation of the wake in the time domain simulations due to computational limits.

\section{Heating Estimates}

To ensure a complete evaluation of the impedance profile of the different beam screen configurations we must produce estimates of the beam-induced heating in the TCTP. This is done by considering the power loss due to the longitudinal impedance of the device. We can consider the beam induced heating in two manners; a worst case scenario at which the frequency of a strong resonance falls on a bunch harmonic, that is we take the maximum possible spectral component for the beam current at that frequency, or we can consider the heating due a broadband impedance convolved with the spectral lines of the beam current, determined by the bunch spacing (typically 20.04MHz for 50ns bunch spacing and 40.04MHz for 25ns bunch spacing in the LHC).

The power loss assuming that the resonance falls upon a beam harmonic is given by

\begin{equation}
P_{loss}  = I_{b}^{2} R_{s} S\left(\omega_{r}\right)
\end{equation}

where $I_{b}$ is the beam current, $R_{s}$ is the shunt impedance of the resonance and $S\left(\omega_{r}\right)$ is the magnitude of the beam power spectrum at the resonant frequency $\omega_{r}$.

The power loss $P_{loss}$ due to a longitudinal impedance $Z_{\parallel}$ in a storage ring can be given by~\cite{metral_cham2012}

\begin{equation}
P_{loss} = \left( f_{rev}eN_{b}n_{bunch}  \right)^{2} \displaystyle\sum\limits_{n = 0}^{\infty} \left( 2 \left| n \lambda \left( \omega_{0} \right)  \right|^{2}  \Re{}e \left( Z_{\parallel} \left( n \omega_{0}\right) \right) \right)
\end{equation}

where $f_{rev}$ is the revolution frequency, $e$ is the electron charge, $N_{b}$ is the bunch population, $n_{bunch}$ the number of bunches in the storage ring $\lambda\left( \omega \right)$ is the bunch current sprectrum is the frequency domain, $\omega_{0} = 2\pi f_{0}$ and $f_{0}~=~\frac{1}{\tau_{b}}$, and $\tau_{b}$ is the bunch spacing.

\begin{figure}
\includegraphics[width=0.45\textwidth]{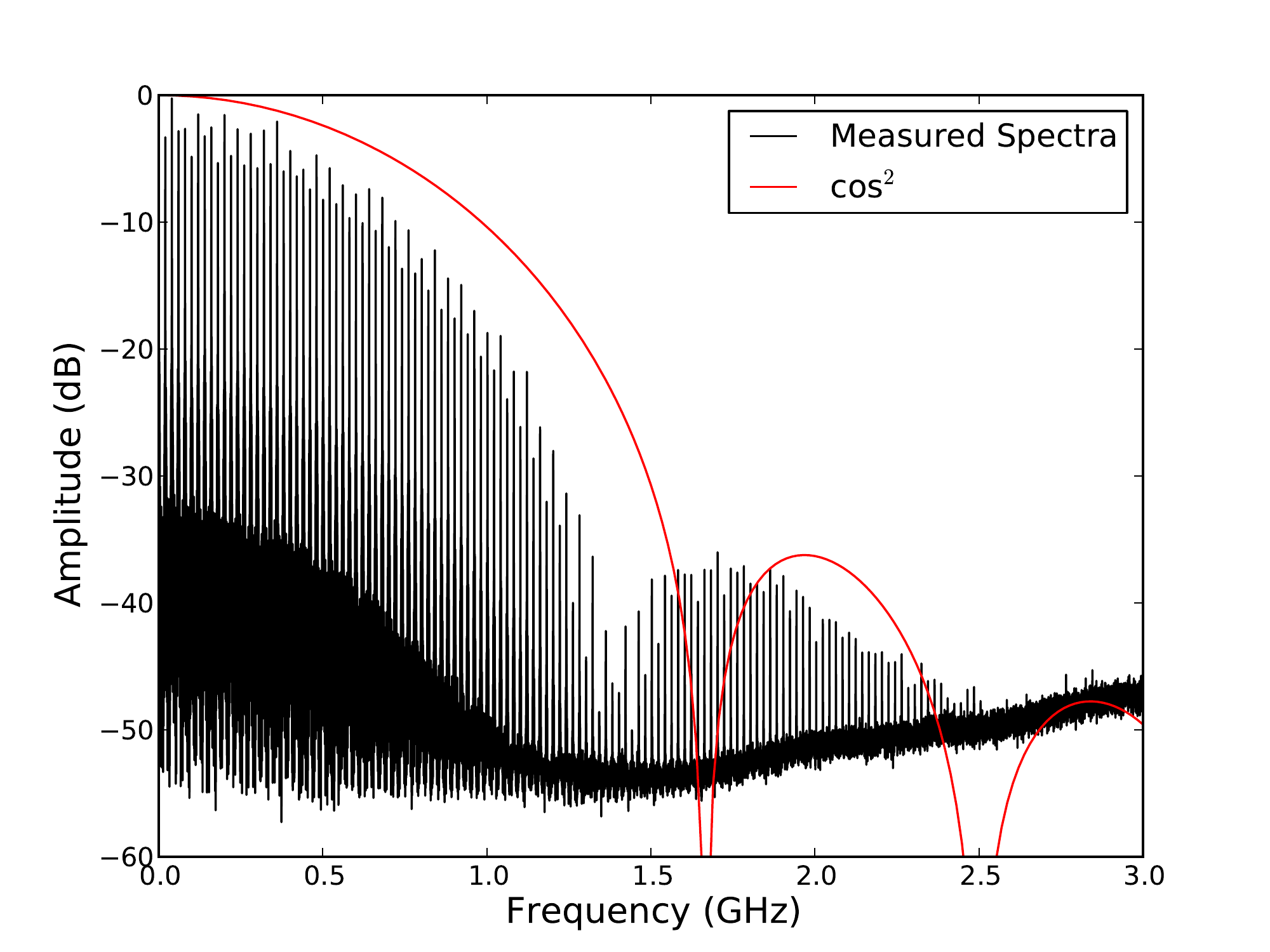}
\label{fig:power_spec}
\caption{Sample measured power spectrum of the LHC beam, in comparison to a cos$^{2}$ distribution. The spectral lines at 20.04MHz intervals due to the 50ns bunch spacing can be seen in the measured spectrum.}
\end{figure}

For heating estimates using the broadband impedance, we use two example beam spectra, one measured and one analytical model, in this case the cos$^{2}$ distribution. The cos$^{2}$ distribution is described in the time domain by the following equations,

\begin{equation}
A\left( t \right) = \int^{\infty}_{-\infty} \lambda \left( \omega \right) e^{j\omega t} d\omega = 
\begin{cases}
cos^{2}\left( \frac{\pi t} {t_{b}} \right) &\textrm{if $| t/2 | \leq t_{b}$}\\
0								&\textrm{if $| t/2 | > t_{b}$}
\end{cases}
\end{equation}

where $t_{b}$ is the length of the bunch in seconds. A sample measured and analytical current spectrum are shown in Fig.~3. Note, the cos$^{2}$ profile has been chosen due its closer matching to the measured spectrum compared to other bunch profiles such as a parabolic or gaussian profile, in particular the presence of the secondary lobe at high frequencies.

For the 50ns LHC-type beam used in 2011, $N_{b}~=~1.45~\times~10^{11}, n_{bunch}~=~1380$, $f_{0,50}~=~20MHz$. 

Estimates for the power loss based on these different methods are given in Table~\ref{tab:power_loss}. It can be seen that the cos$^{2}$ spectrum gives a value of the total power loss two times greater than that using the measured spectrum. This is to be expected due to the higher amplitude at most frequencies (see Fig.~3)

\begin{table}
\begin{center}
\caption{Heating estimates for the TCTP collimator using a measured and an analytical spectrum. These estimates are made assuming a total bunch length of 1.2ns}
\begin{tabular}{| l | c |}
\hline
Distribution & Total Power Loss (W) \\ \hline
Cos$^{2}$  & 17W  \\ \hline
Measured  &  7W \\ \hline
\end{tabular}
\label{tab:power_loss}
\end{center}
\end{table}

Due to the various possible upgrade schemes that may be applied to the LHC, including the movement to 25ns bunch spacing, and future possible development paths such as HL-LHC, we have provided heating estimates for these operation modes.

\begin{table}
\begin{center}
\caption{Beam parameters for different operation modes of the LHC and future upgrades \cite{hl-lhc-para}}
\begin{tabular}{| l | c | c |}
\hline
Operation Mode & $N_{b}$($10^{11}$) & No. of bunches \\ \hline
LHC, $\tau_{b}$=50ns  & 1.45 & 1380 \\ \hline
LHC, $\tau_{b}$=25ns  & 1.15 & 2808 \\ \hline
HL-LHC, $\tau_{b}$=50ns  & 3.3 & 1380 \\ \hline
HL-LHC, $\tau_{b}$=25ns  & 2.2 & 2808 \\ \hline
\end{tabular}
\label{tab:beam_para}
\end{center}
\end{table}

\begin{table}
\begin{center}
\caption{Heating Estimates for the TCTP collimator for different beam modes. $P_{loss}$ is the total power loss and $P_{loss, ferr}$ the power lost in the ferrite tiles.}
\begin{tabular}{| l | c | c | c |}
\hline
Operation Mode & $t_{b}$ (ns) & $P_{loss}$ (W) & $P_{loss, ferr}$ (W)\\ \hline
LHC, $\tau_{b}$=50ns & 1 & 27 & 1 \\ \hline
LHC, $\tau_{b}$=25ns  & 1& 34 & 2 \\ \hline
HL-LHC, $\tau_{b}$=50ns & 1 & 140 & 7 \\ \hline
HL-LHC, $\tau_{b}$=25ns & 1 & 104 & 5 \\ \hline
HL-LHC, $\tau_{b}$=50ns & 0.5 & 374 & 19 \\ \hline
HL-LHC, $\tau_{b}$=25ns & 0.5 & 279 & 14 \\ \hline
\end{tabular}
\label{tab:power_loss_all}
\end{center}
\end{table}

The estimates for these operational modes are given in Table~\ref{tab:power_loss_all}. The beam parameters for the different machine possibilities are given in Table~\ref{tab:beam_para}.

The nominal bunch length for design luminosity is 1ns. There exist two possible bunch lengths for HL-LHC, dependent on operation with and without crab cavities. For operation with crab cavities, the proposed bunch length remains at 1ns, but for operation without crab cavities it is proposed to reduce the bunch length to a minimum of~0.5ns. As can be seen, operation in this mode would lead to a large quantity of power loss into the collimator due to cavity modes, totalling some several hundred watts of power.

Due to the nature of the impedance reduction system it is vital that the ferrites in the TCTP do not heat up to the point that they exceed their Curie temperature $T_{c}$, as at this point they decline in performance as a damping material. For the TCTP we have chosen a ferrite that has a high Curie point ($T_{c} = 375^{\circ}C$ for TT2-111R \cite{tt2-111r-data}), however due to the poor heat transfer in vacuum, it is still vital to know the location of the heat loss as a relatively small power deposition can cause large increases in temperature. By summing the volume and surface losses in the ferrite for all cavity modes, and comparing this to the total power loss in the structure we can obtain an estimate of the total power into the ferrite itself. The amount of power loss into the ferrite is calculated to be between 0-5\% for all cavity modes. In this evaluation we consider the worst case scenario of 5\% for all modes to ensure safe operation of the collimators. These heat loads are summarised in Table~\ref{tab:power_loss_all}.

\section{Summary}

In this paper it has been shown that the use of damping method of impedance mitigation can effectively reduce the beam coupling impedance of cavity modes in a collimator structure. Due to the lack of moving contacts between the parts of the device, it overcomes one of the major disadvantages of RF fingers in moveable devices. In addition, it is shown that a small percentage of the power loss is lost in the damping material itself, meaning that this method can be used for devices exposed to large beam currents (beam current $I_{b}\approx 1A$) and still continue to effectively damp the cavity modes.

\end{document}